\documentclass[floats,floatfix,showpacs,prl,superscriptaddress,twocolumn,nofootinbib]{revtex4-1}

\usepackage[breaklinks]{hyperref}
\usepackage{graphicx, epsfig, amssymb} 
\usepackage{amsmath, amsfonts}
\usepackage{bm} 
\usepackage{color}
\usepackage[usenames]{xcolor}

\newcommand{\non}{\nonumber}
\newcommand{\be}{\begin{equation}}
\newcommand{\ee}{\end{equation}}                  
\newcommand{\bea}{\begin{eqnarray}}
\newcommand{\eea}{\end{eqnarray}}

\renewcommand{\section}[1]{\paragraph{#1. ---}\phantomsection\addcontentsline{toc}{section}{#1}}
\renewcommand{\subsection}[1]{\paragraph*{#1. ---}\phantomsection\addcontentsline{toc}{subsection}{#1}}

\begin{document}

 
\title{Neutron stars in general second order scalar-tensor theory: the case of non-minimal derivative coupling}


\author{Adolfo Cisterna}
\affiliation{Instituto de Ciencias Fis\'icas y Matem\'aticas,\\ Universidad Austral de Chile,\\ Valdivia, Chile}
\email{adolfo.cisterna@uach.cl}
\author{T\'erence Delsate}
\affiliation{Theoretical and Mathematical Physics Dept.\\ University of Mons - UMONS\\ 20, Place du Parc - 7000 Mons - Belgium}
\email{terence.delsate@umons.ac.be}
\author{Massimiliano Rinaldi}
\affiliation{Physics Department, University of Trento,\\  Via Sommarive 14, 38123 Trento, Italy}
\affiliation{INFN - TIFPA, \\ Via Sommarive 14, 38123 Trento, Italy }
\email{massimiliano.rinaldi@unitn.it}

\newcommand{\letter}{paper}


\begin{abstract}

\noindent We consider  the sector of Horndeski's gravity characterized by the 
coupling between the kinetic scalar field term and the Einstein tensor. 
We numerically construct neutron star configurations where the external 
geometry is identical to the Schwarzschild metric but the interior 
structure is considerably different from standard general relativity. We 
constrain the only parameter of this model from the requirement that 
compact configurations exist, and we argue that solutions less 
compact than neutron stars, such as white dwarfs, are also supported.  
Therefore, our model provides an explicit modification of general 
relativity that is astrophysically viable and does not conflict with 
Solar System tests.
\end{abstract}


\maketitle


\section{Introduction}
Compact objects, such as neutron stars are likely to unveil soon an unexplored corner of gravity. Indeed, the upcoming gravitational wave observation facilities are expected to probe the strong field and strong velocity regime of gravity typical of these extreme astrophysical objects \cite{Schutz:1999xj, Berti:2015itd}. To date, the theory of general relativity (GR) has passed all tests with flying colors, therefore it is very difficult to come up with extensions that do not violate any of these tests. On the other hand, modifications of GR are often invoked to cope with quantum effects at high energy and large curvature, in particular in the presence of spacetime singularities.

From a phenomenological point of view  there are many ways to modify gravity and astrophysical data proved useful to assess their viability \cite{Berti:2015itd, Clifton:2011jh}. Historically, one of the first extension of GR consists in the addition of a new dynamical degree of freedom in the form of a scalar field minimally coupled to gravity \cite{Brans:1961sx}. After this pioneering work, the so-called tensor-scalar theories of gravity developed enormously \cite{ScalarTensorReview}.

In this \letter, we consider a particular extension of GR that was originally discovered by Horndeski  in the early seventies \cite{Horndeski:1974wa} consisting in the most general scalar-tensor theory in four dimensions with at most second order differential equations of motion. This feature is particularly desirable since, while several theories with higher derivative terms exist (see e.g. \cite{Alexander:2009tp}), they usually suffer from ghost instabilities or fail to be hyperbolic \cite{Delsate:2014hba} (see also \cite{Biswas:2011ar} for alternative models of gravity without ghosts).

Recently, Horndeski's gravity was rediscovered and called Galileon gravity, the name originating from the invariance of the equations of motion under arbitrary shifts of the scalar field in Minkowski spacetime \cite{Nicolis:2008in}. In fact, the covariantized version of this theory was proven to be equivalent to Horndeski's theory in \cite{Kobayashi:2011nu}. An important subclass of Horndeski's gravity shows self-tuning properties offering a natural explanation for the cosmological constant. This subclass, named ``Fab Four'' in \cite{Charmousis:2011bf}, consists in the combination of four main actions, each with different properties that are in turn relevant for cosmology or astrophysics. 

One particular combination of these actions, named ``John + George'', has the form
\begin{align}
S= &\int\sqrt{-g}d^4x\bigl[ \kappa (R - 2\Lambda) \nonumber\\
&-\frac{1}{2} (\alpha g^{\mu\nu} -\eta  G^{\mu\nu})\nabla_\mu \phi \nabla_\nu \phi\bigr]+S_{m}\,,
\label{eq:action}
\end{align}
where $\phi$ is a real scalar field, $G_{\mu\nu}$ is the Einstein tensor, $\kappa=1/(16\pi G)$, being $G$ is the Newton's constant, $\alpha, \eta$ are real parameters, $\Lambda$ is the cosmological constant, and $S_{m}$ is the action for ordinary matter fields, supposed to be  minimally coupled to gravity in the usual way.
As we will briefly review below, this model has  exact analytical black hole solutions, which are asymptotically anti-de Sitter and have a non-trivial scalar field distribution. Therefore, in order to be a viable alternative to GR, the deviations from asymptotic flatness needs to be small enough to pass at least Solar System tests. A first post-Newtonian analysis of  this model (with vanishing cosmological constant) was presented in \cite{Bruneton:2012zk}, where very stringent constraints on the parameters of the theory were found.

Here, we are interested in spherically symmetric solutions of this theory in the presence of matter, focussing in particular on neutron star models. Our goal is to investigate the effect of the derivative coupling on the structure of neutron stars and to constrain the parameter space. 

While the lack of asymptotic flatness seems to make these models physically unrealistic, it turns out that there exist a region of the parameter space such that the exterior metric is exactly the same as the Schwarzschild one because the scalar field does not backreact on the geometry. On the opposite, inside the star the scalar field behaves as a sort of ``internal hair'' that interacts with the metric and modifies   pressure and matter density distributions considerably with respect to  GR. The size of these effects depends on one parameter and the interesting aspect is that the deviations from GR persist also when this parameter vanishes. Therefore, at the level of the equations of motion, there is no smooth limit to GR.

We begin by reviewing the black hole solutions of  \eqref{eq:action} and their properties. We then consider the modifications induced by the presence of a matter fluid and study the configuration with external Schwarzschild metric. Next, we solve the equations numerically  in the range of mass and density typical of neutron stars and compare the results with GR. 

\section{Black hole solutions}

The equations of motion deriving from the action \eqref{eq:action} read
\begin{eqnarray}
&&G_{\mu\nu}+\Lambda g_{\mu\nu}-H_{\mu\nu}=T_{\mu\nu}^{(m)}\,, \label{eqmetric}\\
&&\nabla_{\mu} J^\mu  =0\,, \label{eqphi}\\
&&\nabla_{\mu}T^{(m)\mu\nu}=0,
\end{eqnarray}
where
\begin{eqnarray}
H_{\mu\nu}&=&\frac{\alpha}{2\kappa}\Big[\nabla_{\mu}\phi\nabla_{\nu}\phi-\frac{1}{2}g_{\mu\nu}\nabla_{\lambda}\phi\nabla^{\lambda}\phi\Big]\nonumber\\
&&+\frac{\eta}{2\kappa}\Big[\frac{1}{2}\nabla_{\mu}\phi\nabla_{\nu}\phi R-2\nabla_{\lambda}\phi\nabla_{(\mu}\phi R_{\nu)}^{\lambda}\nonumber\\%
&&-\nabla^{\lambda}\phi\nabla^{\rho}\phi R_{\mu\lambda\nu\rho}-(\nabla_{\mu}\nabla^{\lambda}\phi)(\nabla_{\nu}\nabla_{\lambda}\phi)\nonumber\\
&&+\frac{1}{2}g_{\mu\nu}(\nabla^{\lambda}\nabla^{\rho}\phi)(\nabla_{\lambda}\nabla_{\rho}\phi)-\frac{1}{2}g_{\mu\nu}(\square\phi)^{2}\nonumber\\
&&+(\nabla_{\mu}\nabla_{\nu}\phi)\square\phi+\frac{1}{2}G_{\mu\nu}(\nabla\phi)^{2}\nonumber\\ 
&&+g_{\mu\nu}\nabla_{\lambda}\phi\nabla_{\rho}\phi R^{\lambda\rho}\Big] \,,\\
T_{\mu\nu}^{(m)}  &=&(\rho + P) u_\mu u_\nu + P g_{\mu\nu}\,, \\ 
J^\mu &=&  \left(  \alpha g^{\mu\nu}-\eta G^{\mu\nu}\right)\nabla_{\nu}\phi\,,
\end{eqnarray}
and where $u$ is the unit 4-velocity of a perfect fluid with pressure $P$ and density $\rho$.

One of the first analytical black hole solutions was found in \cite{Rinaldi:2012vy}, by imposing the diagonal metric 
\be
ds^2 = -b(r) dt^2 + \frac{dr^2}{f(r)} + r^2 d\Omega^2\,,
\label{eq:ds2}
\ee
with  $T_{\mu\nu}^{(m) }= 0$ and $\Lambda=0$.   Thanks to the shift symmetry $\phi \rightarrow \phi + {\rm const}$, the equation of motion for the scalar field turns into the current conservation law \eqref{eqphi}. This symmetry also implies that the solution depends only on $\phi'^{2}$, where, from now on, the prime denotes the derivative with respect to $r$. In addition, to avoid a diverging current norm on the horizon, it is necessary to set $J^{r}=0$ \cite{Charmousis:2015aya}. In this configuration, $\phi'^{2}$ becomes negative outside the event horizon, but this seems not to introduce any thermodynamical instability \cite{Rinaldi:2012vy}. This model was further developed in a series of papers, see e.g. \cite{Anabalon:2013oea, Babichev:2013cya, Cisterna:2014nua, Minamitsuji:2013ura, Bravo-Gaete:2013dca, Kobayashi:2014eva, Minamitsuji:2014hha, Kobayashi:2014wsa, Charmousis:2014zaa}. In particular, it was shown in  \cite{Anabalon:2013oea} that the scalar field can be real outside the event horizon provided one adds a negative cosmological constant $\Lambda$ in the action. In this case, $\phi$ is finite at the horizon but $\phi'$ diverges. The stability and the quasi-normal mode spectrum of these asymptotically $AdS$ solutions were investigated in \cite{Kobayashi:2014wsa, Minamitsuji:2014hha}.

A more general solution was proposed in  \cite{Babichev:2013cya}, where the scalar field has the form $\phi(t,r)=Qt+F(r)$, for some function $F$ and constant $Q$. 
In this case, the equations do not depend explicitly on time and $J^r = 0$ is required to satisfy the $(tr)$ metric equations. Note that this sector of  Horndeski's gravity is equivalent to a $f(R,T,T_{\mu\nu}R^{\mu\nu})$ model, where $T_{\mu\nu}$ is the scalar field stress tensor \cite{Momeni:2014uwa}.

It is interesting to understand in which limit one recovers GR. In the case $Q=0$, this limit is $\Lambda=0$ and $\eta\rightarrow \infty$, revealing that $\eta$ is a non-perturbative parameter if $\alpha$ is fixed \cite{Rinaldi:2012vy}. For $Q\neq 0$,  the limit $\Lambda=\alpha=0$ leads to a non-trivial scalar field with a Schwarzschild metric. This happens because the scalar field satisfies  $H_{\mu\nu} = 0$ so it doesn't backreact on the metric. Such solutions were dubbed ``stealth'' configurations in \cite{Babichev:2013cya}.  

The crucial point is that these considerations hold true only in vacuum. In fact, inside matter $H_{\mu\nu} \neq 0$ and the metric is affected, also when $\Lambda=\alpha=0$, resulting a modified theory of gravity whose deviations from GR are well hidden inside the star, so that they do not conflict with Solar System tests. In the following we focus on this case.

\section{Modified Tolman-Oppenheimer-Volkoff equations}

\begin{figure}
 \includegraphics[scale=.67]{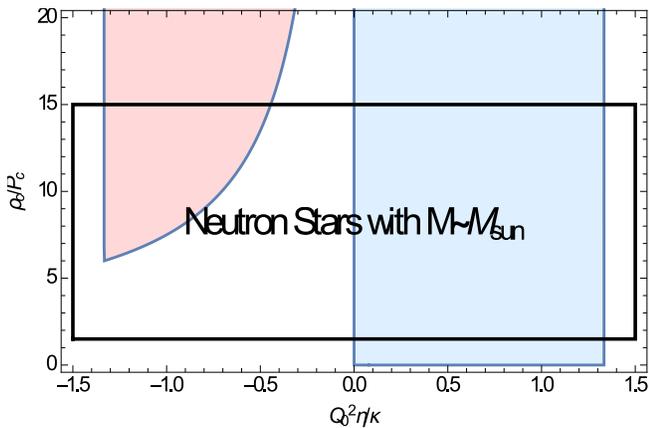}
\caption{Range of $Q_{0}^{2}\eta$ where $F'$ is real and where compact solutions exist. Here, $\rho_c/P_c$ is dimensionless. The red (blue) shaded area corresponds to $F'^2>0$ for $\eta<0$  $(\eta>0)$. Typical solar mass neutron stars lie inside the black rectangle.}
 \label{fig:scF}
 \end{figure}

According to the discussion above, we set $\alpha=\Lambda=0$ and look for neutron star-like solutions assuming the usual equation for the matter fluid
\be
P' + (P + \rho)\frac{b'}{2b}=0\,.
\label{eq:EP}
\ee
The scalar current components are
\bea
J^{t}&=&{\eta Q\over r^{2}b}\left(rf'+b-1\right)\,,\ 
J^{r}={\eta f F'\over r^{2}b}\left[(1-f)b-rfb'\right]\,.\nonumber
\eea
The $(tt)$ and $(rr)$ components of eq.\ \eqref{eqmetric} yield other two equations and we have 
\be
\label{eqf}
rfb' =(1-f)b\,,\ Af' =-B\,,
\ee
where
\bea
A&=&rbf^{-1}(Pr^{2}+4\kappa)-3\eta Q^{2}r\,,\\\non
B&=&3(1-f)\eta Q^{2}\\\non
&+&bf^{-1}[6r^{2}fP+(1+f)r^{2}\rho-4\kappa(1-f)]\,.
\eea
Finally, the  scalar field  equation
\be\label{eqF}
\eta fb F'^{2}=(1-f)\eta Q^{2}+bPr^{2}\,,
\ee
closes the system.
By expanding around $r=0$ (with the regularity conditions  $b(0)=b_0>0,\ b'(0)=0,\ f(0)=1$ and $P(0)=P_{c}$), we find that
\bea\label{sersol}
&&F'^2=r^2 \left(\frac{P_c}{\eta }-\frac{2 Q_{0}^2 (3 P_c+\rho_c)}{3  \left(3   Q_{0}^2\eta-4 \kappa \right)}\right)+{\cal O}(r^{4}),\\\nonumber
&&P = P_c + \frac{(P_c+\rho_c) (3 P_c+\rho_c)}{6 \left(3   Q_{0}^2\eta-4 \kappa \right)}r^2+{\cal O}(r^{4}),
\eea
where we rescaled $Q_{0}^{2}=Q^{2}/b_{0}$. We note that $3Q_{0}^2\eta < {4\kappa}$ corresponds to the condition $P''(0)<0$.  In fact, the latter is the only physically acceptable condition. One can show that, if this condition is violated, $P$ is a monotonic growing function of $r$ so compact configurations cannot exist. This constraint, however, is not sufficient to guarantee that $F'(r)$ is real inside the star when $\eta$ is negative. In fact, in this case one needs also
\be
\frac{1}{4\pi \left( \frac{2\rho_{c}}{3P_{c}}-1 \right)}<Q_0^2|\eta|<\frac{1}{12\pi }\,.
\ee
All these conditions are summarized in Fig.\ \ref{fig:scF}.

For the numerical computations we choose the polytrope equation of state $P = K \rho_B^{1+1/n}$, where $\rho_B$ is the baryonic mass density, $n$ is the polytropic index and $K$ is a constant. Integrating the first law of thermodynamics leads to $\rho = P + (P/K)^{n/(n+1)}$, and we set $K = 123 M_{\odot}^2,\ n=2$ since this models leads to compact objects with accepted mass and radius of neutron stars \cite{Lattimer:2000nx}.  We leave a systematic study of tabulated equation of states (as well as cases with non-vanishing $\alpha$ and $\Lambda$) for future work.

Equipped with the equation of state, we numerically solve \eqref{eq:EP}-\eqref{eqF}, following this strategy: we choose $P(0) = P_c$ for a set of values for $Q$ and integrate the system as a Cauchy problem up to the star's surface at $r= r_*$, determined by $P(r_*)=0$.  Next, we match the interior solution to the Schwarzschild external solution and to a constant scalar field. The boundary conditions are $
b(0) = 1,\ b'(0)=0,\ P(0)=P_c$. Finally, we set $\eta=\pm 1$ without loss of generality since the system depends on the product $Q_{0}^2\eta$ only.

The total mass $M$ is determined by solving $b(r_*) = b_\infty (1-2M/r_*),\ b'(r_*)=2Mb_\infty/r_*^2$, where $M$ is the mass and $b_\infty$ is a constant. We redefine the time coordinate according to $t_p= t \sqrt{b_\infty}$, so  it matches with the time measured by a distant observer in the flat asymptotic region. As a consequence, since the scalar field is linear in time, we must also rescale $Q_\infty = Q/\sqrt{b_\infty}\,$, which is the value measured by the same distant observer. In summary, the input parameters are the central pressure and the bare value of $Q$, from which we numerically compute $Q_{\infty}$ and the mass-radius relation. We work in units where $G=c=1$.

Note that the typical values of central density and pressure for white dwarfs ( $\rho_c\approx 10^6 \ \mbox{g}. \mbox{cm}^{-3}$ and $P_c\approx2.5\ 10^{24}\ \mbox{Dyne}.\mbox{cm}^{-2} $, obtained from realistic equation of state tables describing nuclear matter in the low density regime) lead to larger values of $\rho_c/P_c$ than for neutron stars. Since they are within the range of parameters leading to compact configurations with a real scalar field, we infer that white dwarfs also exists in our theory.

\section{Results}

\begin{figure}
\centering 
\includegraphics[scale=.67]{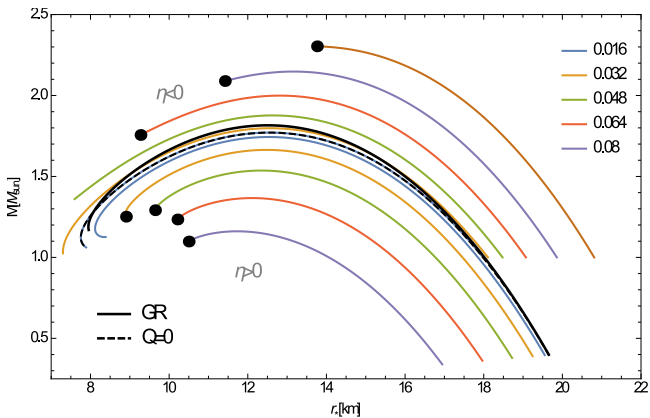}
\caption{The mass-radius relation for  various values of $Q_\infty$ and $\eta = +1,\ -1$. The thick black curve is the  GR prediction while the dashed one has $Q_\infty=0$.  The black dots are the points where the solutions cease to exist. Note that, for $\eta>0$, the curves with $Q_\infty=0.016$ and  $Q_\infty=0.032$ do not reach such a point in the chosen range of central pressures.}
\label{fig:RMPol2}
\end{figure}

\begin{figure}
\centering 
\includegraphics[scale=.67]{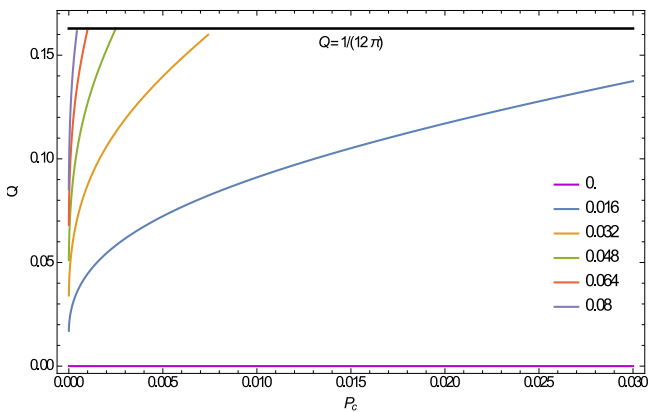}
\caption{Lines of constant $Q_\infty$ in the $(P_c,Q)$ plan. The legend indicates the value of $Q_\infty$. The scalar field is no longer real for $Q>1/(12\pi )$. All curves, except for $Q_\infty=0$, reach a maximal value.}
\label{fig:Qb}
\end{figure}

One of our main results is that, for $\eta<0$ ($\eta>0$), the model allows for bigger (smaller) and more (less) massive neutron stars than in GR. This is illustrated in Fig. \ref{fig:RMPol2}, where we show the mass-radius relation for different values of $Q_\infty$ and $\eta = \pm 1$. Note that a similar effect was found for anisotropic stars in \cite{Silva:2014fca}. From  Fig.\ \ref{fig:Qb} we see that the range of allowed central pressures shrinks with increasing $Q_\infty$. For a given value of $Q_\infty$, the maximum central pressure is determined by the constraint $\eta Q_\infty^{2}<4\kappa /3$ and its value is marked by the black dots in Fig. \ref{fig:RMPol2}, which mark a sharp endpoint for most curves.

We remark that our solutions are not smoothly connected to GR. The limit $Q_\infty=0$ does not lead to general relativity, neither does the $\eta\rightarrow\infty$ limit. This was expected from the discussion in the last section. From the action it would appear that the natural GR limit is $\eta \rightarrow0$, but from eq.\ \eqref{eqF}, we see that the scalar field diverges in this case. 

\begin{figure}
\centering 
\includegraphics[scale=.49]{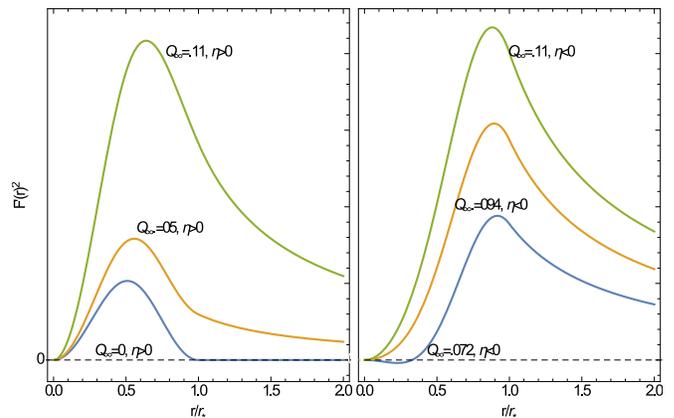}
\caption{Plot of $F'^{2}$ for $\eta>0$ (left) and $\eta<0$ (right). In the case $\eta<0$, $F'$ is imaginary close to the origin for small values of $Q_\infty^2$, and becomes real for larger values of $r$, while it is always real for $\eta>0$.}
\label{fig:ScF}
\end{figure}

It is interesting to look at the scalar field inside the star. As mentioned above, for $Q=0$, $F'$ vanishes exactly outside the star. On the opposite, for $Q\neq 0$, the field smoothy connects, through the star's surface, to the external field that decays at infinity, without affecting the metric. This is shown in Fig.\ \ref{fig:ScF}, where we display the plot of $F'^{2}(r)$. We note that, as anticipated, the field is always real for $\eta>0$, while it can be imaginary for $\eta<0$. In both cases, however, the field is real for $r_{*}<r<\infty$.

We note that where the field becomes imaginary, its kinetic term has the wrong sign in the action. In GR this would inevitably lead to classical and quantum instabilities. However, in the present case, the situation is less clear due to the coupling of the kinetic term to the Einstein tensor. The stability of these configurations with imaginary field is an interesting and open question.

\section{Conclusion}

In this \letter, we investigated the effect of a scalar field with a non-minimal kinetic coupling to gravity. This model is a sector of the Horndeski's theory of gravity and has a deep connection with Galileon gravity. We focused on the case with vanishing cosmological constant and vanishing standard kinetic term coupling, since it admits exact vacuum Schwarzschild solutions. When the scalar field is linear in time we find that neutron stars exist for some range of the only free parameter of the theory $\eta Q^{2}$. The equation of state used here captures very well the main effects of the derivative coupling. In particular, the reality conditions on the scalar field are independent of the equation of state so we do not expect large deviations if one chooses more realistic ones.

We derived a constraint on  $\eta Q^{2}$ from the existence of compact configurations, and argued that less dense stars, such as white dwarfs, are supported by this model. To date, this is the first constraint and astrophysical viability check of this model. The parameter $Q$ is clearly non-perturbative since the ``extremal'' case $Q=0$ still shows a deviation from GR. In general, when $Q\neq 0$ the scalar field grows linearly in time but, since the dynamics depends only on its gradient, there are no diverging physical observables. In fact, the scalar field plays the mere role of a ``clock'', just like the scale factor in cosmology. 

A very attractive feature of our model is that the external metric is the same as in GR. From an observational point of view, this means that Horndeski's stars pass all Solar System tests. In this respect, the model is similar to Palatini $f(R)$ or Eddington-inspired Born Infeld  models \cite{banados:2010} (with similar effects on compact configurations \cite{Pani:2011mg, Pani:2012qb}). In fact, these theories are also indistinguishable from GR in vacuum, but they are modified in the presence of matter fields \cite{Delsate:2012ky}. 

In summary, we show that ours is a sound astrophysical model, which requires a minimal modification of GR and that does not violate any Solar System test. The underlying Horndeski's gravity is currently under thorough scrutiny as a possible candidate to explain fundamental problems, such as inflation and dark energy. Therefore,  we think that also the Horndeski's star deserves full attention as it can shed some light on the general theory. We think that, to further asses the viability of our model, it would be very interesting to address questions such as the the gravitational wave spectrum during a merger of Horndeski's neutron stars or the observational signature on the pulsar emission in the presence of strong magnetic fields, and we hope to report soon on these issues.

\begin{acknowledgments}
\section{Acknowledgement}
We thank P.\ Pani, H.\ Okada da Silva and J.\ Oliva for valuable suggestions. T.D.\ gratefully acknowledges partial support from âNewCompStarâ COST Action MP1304. A.C.\ acknowledges the University of Mons for kind support during the early stage of this project. A.C. work is supported by FONDECYT project N\textordmasculine3150157.  
\end{acknowledgments}

\bibliography{horstar}

\end{document}